\documentclass[prl,twocolumn,superscriptaddress,showpacs]{revtex4}

\makeatletter

\begin{document}

\title{Magnetic-field generation in helical turbulence}
\author{Stanislav Boldyrev}
\affiliation{Department of Astronomy and Astrophysics, University of Chicago, Chicago, IL 60637}
\author{Fausto Cattaneo}
\affiliation{Department of Mathematics, University of Chicago, Chicago, IL 60637} 
\author{Robert Rosner}
\affiliation{Department of Astronomy and Astrophysics, University of Chicago, Chicago, IL 60637}

\date{\today}
\input psfig.sty

\begin{abstract}
We investigate analytically the amplification 
of a weak magnetic field in a homogeneous and isotropic 
turbulent flow lacking reflectional symmetry (helical turbulence). 
We propose that the spectral distributions of magnetic energy and magnetic 
helicity can be found as eigenmodes of a self-adjoint, Schr\"odinger-type 
system of evolution equations. We argue that large-scale and small-scale 
magnetic fluctuations cannot be effectively separated, 
and that the conventional $\alpha$-model is, in general, not an adequate  
description of the large-scale dynamo mechanism. As a consequence,  
the correct numerical modeling of such processes  
should resolve magnetic fluctuations down to the very small,  
resistive scales. 
\pacs{52.30.Cv, 95.30.Qd}
\end{abstract}

\maketitle

{\bf 1.} {\em Introduction.} 
It is well established both analytically and numerically that 
a weak magnetic field can be amplified by the random motions of 
a highly conducting fluid~\cite{cowling,landau,moffatt}. This occurs because  
magnetic field lines 
are generically stretched by 
the random motions of the fluid in which they are (almost) ``frozen''. 
Such mechanisms of turbulent dynamo action are invoked to 
explain the origin of  magnetic fields in astrophysical 
systems, such as planets, stars, 
the interstellar and the intergalactic medium, etc. 

In many cases 
magnetic fields are observed to be strong and ordered on scales much 
larger than the velocity correlation length. The traditional 
view being that the origin of these fields can 
still be explained in the framework of isotropic and homogeneous 
turbulence provided the latter lacks reflectional symmetry.
For this case the helicity integral of the velocity    
can be non-zero, i.e., ${\cal H}=\int {\bf v}\cdot (\nabla \times {\bf v})dV\neq 0$. 

To illustrate this idea let us {\em assume} that the magnetic 
field, compared to the velocity correlation length,~$l_{vel}$, possesses only large scale and small scale components.
The magnetic field evolution is described by the induction equation 
\begin{eqnarray}
\partial_t {\bf B}=\nabla \times ({\bf v}\times {\bf B})+\eta \Delta {\bf B},
\label{induction}
\end{eqnarray}
where $\eta$ is the (collisional) diffusivity. Averaging over the small-scale 
fluctuations,~$l \lesssim l_{vel}$ we obtain the equation for 
the large-scale, or the mean magnetic field~$\bar{\bf B}(x,t)$. 
It can be written in the general form~\cite{moffatt,vainshtein}, 
\begin{eqnarray}
\partial_t \bar{\bf B}=
\nabla \times (\alpha \bar{\bf B} )
+\beta \Delta \bar{\bf B} ,
\label{meanfield}
\end{eqnarray}
where it is assumed that the mean field varies slowly in space, 
and, therefore, its higher-order spatial derivatives can be neglected.  
The parameters $\alpha$ and $\beta$ 
can be estimated on dimensional grounds to be  
$\alpha \sim \langle {\bf v}\cdot (\nabla \times {\bf v})\rangle \tau_{v}$, 
and $\beta \sim \langle {\bf v}^2 \rangle \tau_{v}$, where $\tau_v$ is 
the velocity correlation time. 

In Fourier space,  the linear equation~(\ref{meanfield}) has the 
eigenvalues, $\lambda_1=-\beta k^2$, and 
$\lambda_{2,3}=-\beta k^2 \pm \alpha k$, where $k$ is the wavenumber. 
Thus a  growing eigenmode always exists provided small enough wave 
numbers are allowed in the  system under consideration (galaxy, 
laboratory device, simulation box).
The maximal growth rate is then given by~$\gamma_0=\alpha^2/(4\beta)$, 
and the corresponding scale of the growing mean magnetic field 
is~$l_0\sim 2\beta/\alpha$. In order to comply with the underlying assumption of scale separation it is assumed that 
this scale is much larger than the velocity correlation scale~$l_{vel}$.

The mean-field 
growth rate,~$\gamma_0$  
vanishes if the velocity fluctuations 
possess no helicity,~${\cal H}=0$. 
One therefore might expect that the generation of magnetic fields 
at large scales,~$l > l_{vel}$,  in homogeneous and isotropic 
turbulence may only be possible if the velocity field lacks 
reflectional symmetry, and that such magnetic fields are described 
by the mean-field equation~(\ref{meanfield}). 
This effect is 
traditionally called the $\alpha$-dynamo mechanism. 

However, numerical results suggest that the large-scale magnetic 
field evolution in helical turbulence may not be adequately described 
by the $\alpha$-mechanism~(\ref{meanfield}). For example, 
Vainshtein and Cattaneo~\cite{cattaneo} noted  that the small-scale magnetic 
fields are amplified more effectively than the large-scale ones, 
and when their energy is large 
enough to affect the velocity dynamics, the $\alpha$-mechanism 
may become much less effective. The influence of small-scale magnetic fields  
on the large-scale dynamo mechanism~(\ref{meanfield}) has been stressed 
in many 
works (see, e.g.,~\cite{diamond,brandenburg,blackman}). 

Previous investigations of inconsistencies related 
to the $\alpha$-dynamo mechanism~(\ref{meanfield}) essentially 
concentrated on the nonlinear effects related to dynamo saturation, which,  so far, have resisted exact analytical 
treatment. Present-day direct 
numerical simulations of turbulent dynamo action cannot 
provide conclusive results either, due to quite limited 
numerical resolution, e.g.,~\cite{schekochihin-etal-2005}.

In the present paper we propose that some of the essential physics 
of large-scale magnetic field generation is, in fact,  captured 
already at the initial, {\em kinematic} stage of dynamo action.  
Our analysis is based on the exactly solvable model of 
dynamo action due to Kazantsev \cite{kazantsev}. Despite the many simplifying assumptions about the statistics of the 
velocity field this model has proven to be a valuable tool in understanding the dynamo mechanism.
In particular, it treats the induction equation~(\ref{induction}) exactly, and
it allows a rigorous derivation of the $\alpha$-model equation~(\ref{meanfield}).  So far 
only the non-helical case has been extensively analyzed in the 
literature, e.g.,~\cite{kazantsev,vainshtein-kichatinov,schekochihin}. 
Here we address the problem in its full generality.

As an important new result, we show that the evolution equations for the magnetic-energy and 
the magnetic-helicity have self-adjoint structure (we note here that 
although the kinematic-dynamo equations have been known for over 20 years, 
their self-adjoint structure had not so far been discovered). As a 
consequence, in the kinematic regime, the spectrum of magnetic 
fluctuations can be expressed as a sum of eigenfunctions of a 
Schr\"odinger-type equation with imaginary time, where the 
eigenvalue, $\lambda$, gives the growth rate of the 
corresponding mode. 

In analogy to the quantum-mechanical states in a 
potential well,  the eigenmodes growing with $\lambda \leq 2\gamma_0$ 
correspond to ``travelling particles,'' i.e. they are correlated at 
the system size. By contrast, the faster-growing modes 
(with $\lambda >2\gamma_0$) correspond to ``trapped particles''; their  
correlation lengths are less than infinity, and they fill 
the whole range of scales down to the resistive scale. 
At any given scale the modes with $\lambda > 2\gamma_0$ may rapidly become 
dominant over the slowly growing non-localized modes. 

The eigenmodes with $\lambda > 2\gamma_0$ are not captured by 
the mean-field equation~(\ref{meanfield}), consequently  
the $\alpha$-dynamo model~(\ref{meanfield}), 
based on the assumption of scale separation and on small-scale smoothing is inadequate. To describe 
the large-scale dynamo mechanism correctly, 
numerical simulations of uniform, isotropic, helical turbulence 
must resolve the full range of scales from $l_0$ to 
to~$l_{\eta}$. Furthermore, the origin of large-scale fields, such as those observed in astrophysical situations,
may be related to the nonzero large-scale average of the fluctuating part of the field, and not to the mean-field as 
described by mean-field models.

{\bf 2}. {\em The Kazantsev model for helical kinematic dynamo.}
Kazantsev \cite{kazantsev} and Kraichnan~\cite{kraichnan} 
introduced the solvable models in the theory of passive random 
advection. The essential assumption  is that the 
random velocity field is Gaussian 
and short-time correlated. It is also assumed that  
the velocity field has zero mean, 
$\langle {\bf v} \rangle=0$, so that 
the problem is completely specified by 
the velocity covariance tensor. For the 
statistically homogeneous and isotropic case, the 
covariance can be written as 
\begin{eqnarray}
\langle {v^i}({\bf x},t){v^j}({\bf x}',t') \rangle =
\kappa^{ij}(|{\bf x}-{\bf x}'|)\delta(t-t'), 
\label{corr}
\end{eqnarray}
where $\kappa^{ij}$ is an isotropic tensor.
For mirror-symmetric velocities, the correlation tensor $\kappa^{ij}$ 
is symmetric with 
respect to the interchange of the indices $i$ and~$j$. In the general case, however,
this tensor has both symmetric and antisymmetric parts,
\begin{eqnarray}
\kappa^{ij}(x)=\kappa_N 
\left(\delta^{ij}-\frac{x^ix^j}{x^2}\right)+\kappa_L \frac{x^ix^j}{x^2}+g\epsilon^{ijk}x^{k}.  
\label{kappa}
\end{eqnarray}
The first two terms at the right-hand side of (\ref{kappa}) represent the symmetric, 
non-helical part, 
while the function $g(x)$ describes the helical part of the velocity correlation tensor. 
Here 
$\epsilon^{ijk}$ is the completely anti-symmetric 
pseudo-tensor, and summation over the repeated indices is assumed.  
The requirement that  the velocity be incompressible implies that $\kappa_N(x)=\kappa_L(x)+x \kappa^{\prime}_L(x)/2$, 
where the primes denotes  derivatives with respect to~$x$.

The magnetic field correlator can similarly be introduced:
$H^{ij}(x,t)=\langle B^i({\bf x}, t)B^j(0,t)\rangle$, satisfying
\begin{eqnarray}
H^{ij}=M_N\left(\delta^{ij}-\frac{x^ix^j}{x^2} \right)+M_L\frac{x^ix^j}{x^2}
+K\epsilon^{ijk}x^k,
\label{bcorr}
\end{eqnarray}
where the corresponding solenoidality constraint 
implies $M_N=M_L+x  M'_L/2 $. Our goal is to find the 
functions $M_L(x, t)$ and $K(x, t)$ that contain the information 
about the magnetic energy and the magnetic helicity. 

Differentiating $H^{ij}(x,t)$ 
with respect to~$t$ and making use of (\ref{induction}), (\ref{corr}), 
and (\ref{kappa}),  we obtain after a cumbersome but straightforward calculation,   
that the magnetic correlation tensor 
obeys 
\begin{eqnarray}
\partial_t H^{ij}={\hat R}^{imn}{\hat R}^{jrt}\left(T^{mr} H^{nt} \right), 
\label{H1}
\end{eqnarray}
where ${\hat R}^{imn}=\epsilon^{ikl}\epsilon^{lmn}\nabla_k$, and   
$\nabla_k\equiv \partial/\partial x^k$. An analogous, although not identical, representation 
of this equation was derived in~\cite{vainshtein-kichatinov}, but see also the derivation 
in~\cite{boldyrev1, schekochihin}. 
The tensor $T^{ij}$ can be represented 
in the following form:
\begin{eqnarray}
T^{ij}=\frac{A}{\sqrt{2}}\left(\delta^{ij}-\frac{x^ix^j}{x^2}\right)+B\frac{x^ix^j}{x^2}
+\frac{C}{\sqrt{2}}\epsilon^{ijk}\frac{x^{k}}{x}, 
\end{eqnarray}
where 
\begin{eqnarray}
A(x)&=&\sqrt{2}\left(\kappa_N(0)-\kappa_N(x)+2\eta\right),\label{A}\\
B(x)&=&\left(\kappa_L(0)-\kappa_L(x)+2\eta\right),\label{B}\\
C(x)&=&\sqrt{2}\left(g(0)-g(x)\right)x,\label{C}
\end{eqnarray}
and symbol $B(x)$ in~(\ref{B}) should not be confused 
with the magnetic field~$B^i(x,t)$ in Eq~(\ref{bcorr}).   
Hereinafter, 
we adopt the notation $\kappa_0\equiv \kappa_L(0)=\kappa_N(0)$, 
and $g_0\equiv g(0)$.  

Equation~(\ref{H1}) can 
be considerably simplified, since the magnetic-field tensor~(\ref{bcorr}) 
contains only two independent functions,~$M_L$ and~$K$. The reduced equations 
were derived in~\cite{vainshtein-kichatinov}, however, the symmetric 
structure of the tensor equation~(\ref{H1}) was not preserved.  
In the next section, we derive the reduced equations, keeping their symmetric structure 
intact. 
In this way, we reveal the self-adjoint nature of the equations 
which allows us to gain new insight 
into the large-scale dynamo mechanism, and to elucidate the 
limitations of the conventional $\alpha$-dynamo paradigm presented in the introduction.

{\bf 3}. {\em The self-adjoint dynamo equations.} 
In this section we show that (\ref{H1}) is self-adjoint. We begin by rewriting 
(\ref{H1}) in the equivalent form,
\begin{eqnarray}
\partial_t H^{ij}={\hat D}^{ij}_{l\tilde l}{J}^{l\tilde l}_{nt} H^{nt},
\label{H2}
\end{eqnarray}
where $\hat D$ is the self-adjoint differential operator 
\begin{eqnarray}
{\hat D}^{ij}_{l\tilde l}=\epsilon^{ikl}\epsilon^{j{\tilde k}{\tilde l}}
\nabla_k \nabla_{\tilde k}, 
\label{D}
\end{eqnarray}
and the matrix $J$ is symmetric,
\begin{eqnarray}
{J}^{l\tilde l}_{nt}= \epsilon^{npl}\epsilon^{tq\tilde l}T^{pq},
\label{J}
\end{eqnarray}
i.e., this matrix does not change under the interchange of its lower and 
upper sets of indices. We now express the operators~$\hat D$ and~$J$ in the basis 
defined by the three orthogonal ``vectors'':  
\begin{eqnarray}
\xi^{ij}_1&=&\frac{1}{x\sqrt{2}}\left(\delta^{ij}-\frac{x^ix^j}{x^2}\right)\label{basis1},\\ 
\xi^{ij}_2&=&\frac{x^ix^j}{x^3}\label{basis2}, \\
\xi^{ij}_3&=&\frac{1}{x\sqrt{2}}\epsilon^{ijs}\frac{x^s}{x}.\label{basis3}
\end{eqnarray}  
This  is possible since the functions $\kappa^{ij}$, and $H^{ij}$ can themselves be expanded in this basis. 
The normalization of the vectors $\xi_1^2=\xi_2^2=\xi_3^2=1/x^2$ is chosen in such a way as 
to preserve the self-adjoint structure of the differential operator, $\hat D$, as we will see 
presently. A straightforward calculation leads to:
\begin{eqnarray}
J=\left[ \begin{array}{ccc}
B & A & 0\\
A & 0 & C\\
0 & C & B
\end{array}\right],
\label{J1}
\end{eqnarray}
\begin{eqnarray}
{\hat D}=\left[ \begin{array}{ccc}
\frac{\partial^2}{\partial x^2}  & 
-\frac{\partial}{\partial x}\frac{\sqrt{2}}{x} & 0 \\
\frac{\sqrt{2}}{x}\frac{\partial }{\partial x} & -\frac{2}{x^2} & 0 \\ 
0 & 0 & \frac{1}{x^2}\frac{\partial}{\partial x}x^4\frac{\partial}{\partial x}\frac{1}{x^2}
\end{array}
\right],
\label{D1}
\end{eqnarray}
where both operators are manifestly self-adjoint. 


We now make the following crucial observation. It can be verified directly 
that the operator~${\hat D}$ can be factorized as ${\hat D}=-{\hat R}{\hat R}^T$, where 
\begin{eqnarray}
{\hat R} =\left[ \begin{array}{ccc}
0 & \frac{\partial}{\partial x} & 0\\
0 & \frac{\sqrt{2}}{x} & 0\\
0 & 0& -\frac{1}{x^2}\frac{\partial }{\partial x}x^2
\end{array}\right]. 
\label{R}
\end{eqnarray} 
This factorization immediately allows  the dynamo equations to be transformed into self-adjoint 
form. Let us introduce the 
vector~$W$ such that ~$H={\hat R}W$. As can be directly checked, with this definition 
the vector~$H$ automatically satisfies the 
solenoidality condition, 
i.e., its components in the basis~(\ref{basis1}-\ref{basis3}) can be represented  
as $H=\{\sqrt{2}xM_N, xM_L, \sqrt{2}x^2K \}$, where 
$M_N=M_L+\frac{1}{2}xM_L'$, and $M_L$ and $K$ are some independent functions 
(compare this with~(\ref{bcorr})). 
We further require that the vector $W$ satisfies the equation:
\begin{eqnarray}
\partial_t W =-{\hat R}^T {J} {\hat R} W,
\label{W}
\end{eqnarray}
then, clearly, the function~$H$ obeys the dynamo equations~(\ref{H2}), as can be verified by applying the 
operator~$\hat R$ 
to both sides of Eq.~(\ref{W}).  The operator in the right-hand-side of (\ref{W}) 
is now explicitly self-adjoint. This representation constitutes the formal solution of our problem. 

For practical purposes, (\ref{W}) can be further simplified 
since only two components of the vector~$H$ are 
independent. Conveniently, the necessary reduction is already present in~(\ref{W}). Indeed, 
calculating the operator in the right-hand-side 
of~(\ref{W}), one sees that it acts only 
on the second and the third components of the vector~$W$, so the system is automatically reduced to the 
two independent equations that preserve the initial symmetry structure. 
The reduced equations have the self-adjoint form 
\begin{eqnarray}
\partial_t W =-{\tilde R}^T {\tilde J} {\tilde R} W,
\label{W2}
\end{eqnarray}
where ${\tilde R}$ is the reduced form of the operator~$\hat R$, and $\tilde J$ is the reduced form of~$J$: 
\begin{eqnarray}
{\tilde R}=\left[ \begin{array}{cc}
\frac{\sqrt{2}}{x} & 0\\
0 & -\frac{1}{x^2}\frac{\partial}{\partial x}x^2
\end{array}\right],\quad    
{\tilde J}=\left[ \begin{array}{cc}
{\hat E} & C\\
C & B
\end{array}\right].
\label{reduced}
\end{eqnarray}
Here we introduced 
the self-adjoint operator 
\begin{eqnarray}
{\hat E}= -\frac{1}{2} x\frac{\partial}{\partial x} B\frac{\partial}{\partial x}x  + \frac{1}{\sqrt{2}}(A-xA').
\label{E} 
\end{eqnarray}
The validity of (\ref{W2}, \ref{reduced}) can be verified most easily  by direct 
calculation of the right-hand sides of~(\ref{W}) and~(\ref{W2}). For convenience, we  write out 
the matrix form of Eq.~(\ref{W2}) explicitly:
\begin{eqnarray}
\left[\begin{array}{c}
{\partial_t} W_2\\
{\partial_t} W_3
\end{array}\right]=
\left[\begin{array}{cc}
-\frac{\sqrt{2}}{x}{\hat E}\frac{\sqrt{2}}{x} & \frac{\sqrt{2}}{x^3}C\frac{\partial}{\partial x} x^2\\
-{x^2}\frac{\partial}{\partial x}C\frac{\sqrt{2}}{x^3} & x^2\frac{\partial }{\partial x}\frac{B}{x^4}
\frac{\partial }{\partial x}{x^2}
\end{array}
\right]
\left[\begin{array}{c}
W_2\\
W_3
 \end{array}\right],
\label{W3}
\end{eqnarray}
and the relation $H={\tilde R}W$ reads:

\begin{eqnarray} 
M_L=\frac{\sqrt{2}}{x^2}W_2, \quad K=-\frac{1}{\sqrt{2}x^4}\frac{\partial}{\partial x}\left(x^2 W_3\right).
\label{change}
\end{eqnarray}
Equations (\ref{W3}) and~(\ref{change}) are the main result of this section. 

For completeness, we note that 
the equations for the functions~$M_L$ 
and~$K$ were first derived by 
Vainshtein \& Kichatinov~\cite{vainshtein-kichatinov} in the   
non self-adjoint form:    
\begin{eqnarray}
\partial_t M=\frac{1}{x^4}\frac{\partial}{\partial x}\left( x^4\kappa 
\frac{\partial M}{\partial x} \right) +GM -4hK,
\label{system1} \\
\partial_t K = \frac{1}{x^4}\frac{\partial}{\partial x}
\left( x^4\frac{\partial}{\partial x}\left(\kappa K+ hM \right) \right).
\label{system2}
\end{eqnarray}
Here we adopt the standard notation 
$\kappa = 2\eta +\kappa_L(0)-\kappa_L(x)$, $h=g(0)-g(x)$,  $ G=\kappa''+4\kappa'/x $, and $M\equiv M_L$. 
These equations also follow from (\ref{W3}). We also note 
that the Fourier-space versions of (\ref{system1}, \ref{system2}) were 
derived  by Kulsrud \& Anderson~\cite{kulsrud} for the limit of 
large magnetic Prandtl number (ratio of fluid viscosity to resistivity), 
and by Berger \& Rosner~\cite{berger} for the general case.

{\bf 4}. {\em Discussion and conclusion}.
In systems with no kinetic helicity 
(i.e., with $C(x)\equiv 0$), magnetic dynamo action is always possible  if the magnetic 
Reynolds number is large enough~\cite{vainshtein-kichatinov,boldyrev}. 
Since the ``helical'' 
terms in (\ref{W3}) have a {\it de}stabilizing effect, systems with kinetic 
helicity should exhibit dynamo action as well. 
A rigorous analysis of the dynamo mechanism requires knowledge of  
the exact spectrum of~(\ref{W3}) which is  
not known for a general velocity correlator, $\kappa^{ij}(x)$. However, the typical 
behavior of the solution can be understood as follows. We introduce the 
mean-field growth rate, $\lambda_0=g^2_0/\kappa_0$. From the 
asymptotic behavior of system~(\ref{W3}) as $x \to \infty$, one  can show that its   
eigenmodes with $\lambda >\lambda_0$ are localized, and the closer the growth 
rate to $\lambda_0$, the larger the correlation length. On the other hand, the eigenmodes 
corresponding to  $\lambda \leq \lambda_0$ have infinite correlation length.     

The  formal analogy between  Eqs.~(\ref{W3}) and  imaginary-time quantum 
mechanic suggests that the eigenfunctions with $\lambda>\lambda_0$ correspond 
to ``particles'' trapped by the potential provided by velocity fluctuations, while  
the eigenfunctions with $\lambda\leq\lambda_0$  correspond to ``travelling particles''. 
In the non-helical case, where only trapped particles have positive 
eigenvalues, the spacing between the eigenvalues decreases with increasing 
magnetic Reynolds number, see e.g., \cite{schekochihin}. It is reasonable 
to expect that the same result holds for the helical case.

We now explain the extent to which the 
mean-field equation~(\ref{meanfield}) describes the dynamo mechanism. 
Remarkably, in the Kazantsev model, equation~(\ref{meanfield}) can be derived 
{\it exactly}, see, e.g., \cite{vainshtein,boldyrev1}, which 
allows one to find its precise relation to Eq.~(\ref{W3}). 
In the derivation, $\bar{\bf B}(x,t)$ is 
the field averaged over the statistical ensemble of the velocity 
fluctuations~(\ref{corr}), and the coefficients 
in the mean-field equation~(\ref{meanfield}) are  
given by $\alpha=g_0$, $\beta=\eta+\kappa_0/2$~\cite{boldyrev1}. In this 
model, $\lambda_0=2\gamma_0$.  
Eq.~(\ref{meanfield}) can therefore be used to derive the evolution 
equation for 
the correlator of the mean field, ${\bar H}^{ij}(x,t)=
\langle \bar{B}^i(x,t)\bar{B}^j(0,t)\rangle$, 
where the brackets denote averaging over the random initial conditions of 
the magnetic field.  

It can be checked that this evolution equation formally coincides with 
the large-scale asymptotic ($x\to \infty$) of our system~(\ref{W3}); 
consequently it can only be used to obtain 
the large-scale asymptotics of the solutions of Eq.~(\ref{W3}).  
More precisely, representing the magnetic field 
as ${B}^i(x,t)={\bar{B}}^i(x,t)+\delta {B}^i(x,t)$, where $\delta {B}^i$  is 
the fluctuating part, we can write $H^{ij}(x,t)=
\langle \bar{B}^i(x,t)\bar{B}^{j}(0,t)\rangle+ \langle \delta{B}^i(x,t)\delta{B}^j(0,t)\rangle$. 
We note that while the system~(\ref{W3}) describes the exact  
function $H^{ij}(x,t)$, the mean-field equation (\ref{meanfield}) captures only 
its slowly growing non-fluctuating part ${\bar H}^{ij}$, which is described by
the large-scale asymptotics of $H^{ij}(x,t)$, 
since the correlation of the fluctuations vanishes for infinite 
scale separation.  


In summary, we have used the Kazantsev model to compare the exact 
spectra of magnetic energy and helicity with the predictions of 
the $\alpha$-model~(\ref{meanfield}). We 
have demonstrated that the large-scale asymptotics ($x\to \infty$) of the exact solution 
is described by the non-localized  
eigenmodes ($\lambda\leq \lambda_0$) of the self-adjoint dynamo equations~(\ref{W3}).  
This asymptotics can also be derived from the mean-field $\alpha$-dynamo 
equation~(\ref{meanfield}). However, model~(\ref{meanfield}) misses the faster 
growing eigenmodes 
with~$\lambda>\lambda_0$, which are present in~(\ref{W3}). The 
correlation lengths of these eigenmodes are generally not small. 
They fill the range of scales from the system scale to the 
the resistive ones, so these modes may not be removed by a small-scale smoothing  
procedure. In numerical simulations or astrophysical 
applications these modes may dominate the slowly 
growing ``mean-field'' modes. The correct description of the dynamo 
mechanism thus  requires the resolution of the whole range of scales 
available to the magnetic field.  

We are grateful to Samuel Vainshtein for many important discussions. 
This work was supported by the NSF Center for Magnetic
Self-Organization in Laboratory and Astrophysical Plasmas
at the University of Chicago.

\end {document}